\newcommand{\be}{\begin{equation}}  
\newcommand{\ee}{\end{equation}}  
\newcommand{\beq}{\begin{eqnarray}}  
\newcommand{\eeq}{\end{eqnarray}}
\documentclass{PoS}

\usepackage{amsmath}
\usepackage{graphicx}
\usepackage{mathtools}

\title{Importance of closed quark loops for lattice QCD studies of tetraquarks}

\ShortTitle{Importance of closed quark loops for lattice QCD studies of tetraquarks}

\author{\speaker{Joshua Berlin}${}^{1}$, Abdou Abdel-Rehim${}^{2}$, Constantia Alexandrou${}^{2,3}$, Mattia Dalla Brida${}^{4}$, Jacob Finkenrath${}^{3}$, Mario Gravina${}^{5}$, Theodoros Leontiou${}^{6}$, Marc Wagner${}^{1}$\\

\email{a.abdel-Rehim@cyi.ac.cy},
\email{alexand@ucy.ac.cy},
\email{berlin@th.physik.uni-frankfurt.de},
\email{mattia.dallabrida@gmail.com},
\email{j.finkenrath@cyi.ac.cy},
\email{mario.gravina@fis.unical.it},
\email{t.leontiou@frederick.ac.cy},
\email{mwagner@th.physik.uni-frankfurt.de}\\

        ${}^{1}$Goethe-Universit\"at Frankfurt am Main, Institut f\"ur theoretische Physik, \\ $\quad$ Max-von-Laue-Stra{\ss}e 1, D-60438 Frankfurt am Main, Germany\\        
        ${}^{2}$Computation-based Science and Technology Research Center, The Cyprus Institute, \\ $\quad$ 20 Kavafi Street, 2121 Nicosia, Cyprus\\
        ${}^{3}$Department of Physics, University of Cyprus, P.O. Box 20537, 1678 Nicosia, Cyprus\\
        ${}^{4}$NIC, DESY, Platanenallee 6, 15738 Zeuthen, Germany\\
		$\quad$ \& Dipartimento di Fisica, Universit\`a di Milano-Bicocca, and INFN,
sezione di Milano-Bicocca, Piazza della Scienza 3, I-20126 Milano, Italy \\
        ${}^{5}$Universit\`a della Calabria, Via Pietro Bucci, 87036 Arcavacata di Rende Cosenza, Italy \\
        ${}^{6}$Department of Mechanical Engineering, Frederick University, 7 Y. Frederickou Str.,\\ 			$\quad$ Pallouriotisa, Nicosia 1036, Cyprus
        }

\abstract{
To investigate the light scalar tetraquark candidate $a_0(980)$ (quantum numbers $J^P=0^+$), a correlation matrix including a variety of two- and four-quark interpolating operators has to be computed. We discuss efficient techniques to compute the elements of this correlation matrix, in particular diagrams with closed quark loops. Furthermore, we present evidence that such diagrams are not negligible given our precision, and their contribution is essential to obtain physically meaningful results. In particular, we find indications of the existence of an "additional" state around the two-particle thresholds of $K + \bar{K}$ and $\eta + \pi$, which could correspond to the $a_0(980)$ meson.
}

\FullConference{34th annual International Symposium on Lattice Field Theory\\
		24-30 July 2016\\
		University of Southampton, UK}

\begin{document}


\section{\label{sec:mot} Motivation}

The mass ordering of the light scalar mesons $\sigma$, $\kappa$, $f_0(980)$ and $a_0(980)$, as observed by experiments, is inverted compared to expectations from a conventional quark-antiquark picture. Therefore, these mesons are frequently discussed as tetraquark candidates. Assuming such a four-quark structure the expected mass ordering is consistent with experimental results and the degeneracy of $f_0(980)$ and $a_0(980)$ is easy to understand.

Several lattice QCD studies of the light scalar mesons have been published in the last couple of years (cf.\ e.g.\ \cite{Bernard:2007qf,Gattringer:2008be,Prelovsek:2008qu,Liu:2008ee,Prelovsek:2010kg,Wakayama:2012ne, Prelovsek:2013ela,Wakayama:2014gpa,Dudek:2016cru}). In this work we continue our investigation of the $a_0(980)$ meson \cite{Daldrop:2012sr,Alexandrou:2012rm,Wagner:2012ay, Wagner:2013nta,Wagner:2013jda,Wagner:2013vaa,Abdel-Rehim:2014zwa,Berlin:2015faa}. Specifically, we show the importance of diagrams with closed quark loops, and present results which support the existence of an additional state around the two-particle thresholds of $K + \bar{K}$ and $\eta+\pi$. This could correspond to the $a_0(980)$ meson.


\section{\label{sec:operators}Interpolating operators}

Our investigation is based on a $6 \times 6$ correlation matrix
\begin{align}
\label{EQN100} C_{j k}(t) = \Big\langle \mathcal{O}^j(t_2) \mathcal{O}^{k \dag}(t_1) \Big\rangle \quad , \quad t = t_2 - t_1 .
\end{align}
The interpolating operators $\mathcal{O}^j$ generate quantum numbers $I(J^P) = 1(0^+)$,
\begin{align}
\mathcal{O}^1 = \mathcal{O}^{q\bar q} =& \sum_{\bf{x}} \Big( {\bar d}_{\bf x} {u}_{\bf x} \Big) \label{eq:operatorone}
 \\
\mathcal{O}^2 = \mathcal{O}^{K \bar{K} \text{, point}} =& \sum_{\bf{x}} \Big( {\bar s}_{\bf x} \gamma_5 {u}_{\bf x} \Big) \Big( {\bar d}_{\bf x} \gamma_5 {s}_{\bf x} \Big) \label{eq:operatortwo}
\\
\mathcal{O}^3 = \mathcal{O}^{\eta_s \pi \text{, point}} =& \sum_{\bf{x}} \Big( {\bar s}_{{\bf x}} \gamma_5 {s}_{{\bf x}} \Big) \Big( {\bar d}_{{\bf x}} \gamma_5 {u}_{{\bf x}} \Big) \label{eq:operatorthree}
 \\
\mathcal{O}^4 = \mathcal{O}^{Q \bar Q} =& \sum_{\bf{x}} \epsilon_{abc} \Big( {\bar s}_{{\bf x}, b} {(C \gamma_5)} {\bar d}^T_{{\bf x}, c} \Big) \epsilon_{ade} \Big( {u}^T_{{\bf x}, d} {(C \gamma_5)} s_{{\bf x}, e} \Big) \label{eq:operatorfour}
\\
\mathcal{O}^5 = \mathcal{O}^{K\bar{K} \text{, 2-part}} =& \sum_{{\bf x,y}} \Big( {\bar s}_{{\bf x}} \gamma_5 {u}_{{\bf x}} \Big) \Big( {\bar d}_{{\bf y}} \gamma_5 {s}_{{\bf y}} \Big) \label{eq:operatorfive}
\\
\mathcal{O}^6 = \mathcal{O}^{\eta_s \pi \text{, 2-part}} =& \sum_{\bf{x,y}} \Big( {\bar s}_{{\bf x}} \gamma_5 {s}_{{\bf x}} \Big) \Big( {\bar d}_{{\bf y}} \gamma_5 {u}_{{\bf y}} \Big) , \label{eq:operatorsix}
\end{align}
where $C$ is the charge conjugation matrix. The operator $\mathcal{O}^{q\bar q}$ generates a standard quark-antiquark state, while all other operators generate four-quark states. $\mathcal{O}^{K \bar{K} \text{, point}}$ and $\mathcal{O}^{\eta_s \pi \text{, point}}$ are of mesonic molecule structure ($K \bar{K}$ and $\eta_s \pi$), while $\mathcal{O}^{Q \bar Q}$ corresponds to a diquark-antidiquark pair (we use the lightest (anti)diquarks with spin structure $C \gamma_5$ \cite{Jaffe:2004ph,Alexandrou:2006cq,Wagner:2011fs}). These three operators are intended to model the expected structures of possibly existing four-quark bound states, i.e.\ of tetraquarks. The remaining two operators $\mathcal{O}^{K\bar{K} \text{, 2-part}}$ and $\mathcal{O}^{\eta_s \pi \text{, 2-part}}$ generate two independent mesons ($K + \bar{K}$ and $\eta_s + \pi$) and, hence, should be suited to resolve low-lying two-particle scattering states.


\section{Lattice setup}

Computations have been performed using 500 Wilson clover gauge link configurations generated with 2+1 dynamical quark flavors by the PACS-CS collaboration \cite{Aoki:2008sm}, with four different source timeslices per configurations. The lattice size is $32^3 \times 64$, with a lattice spacing of $a \approx 0.09 \, \textrm{fm}$ and a $u/d$ quark mass corresponding to $m_\pi \approx 300 \, \textrm{MeV}$.


\section{\label{sec:tecandconcept}Technical aspects}

For each diagram of the correlation matrix \eqref{EQN100} we have implemented various methods of computation (combinations of point-to-all and stochastic timeslice-to-all propagators, the one-end trick and sequential propagators; cf.\ \cite{Abdel-Rehim:2014zwa} for more details). In order to
select for each diagram the most efficient method, we study the ratio $R^{(a),(b)} = \Delta C^{(a)}(t) \sqrt{\tau^{(a)}} / \Delta C^{(b)}(t) \sqrt{\tau^{(b)}}$, which compare statistical errors of the correlation matrix element obtained with method~$(a)$ and $(b)$, weighted by the corresponding computing times. $R^{(a),(b)} > 1$ indicates that method~$(a)$ is superior to method~$(b)$, while $R^{(a),(b)} < 1$ indicates the opposite.

\begin{figure}[htb]
\begin{center}
\hspace{-0.8cm}\includegraphics[width=7.6cm,page=5]{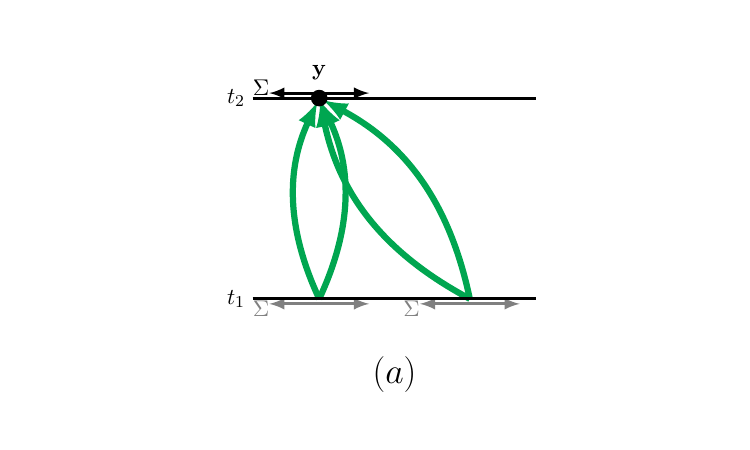}
\hspace{-3.8cm}\includegraphics[width=7.6cm,page=6]{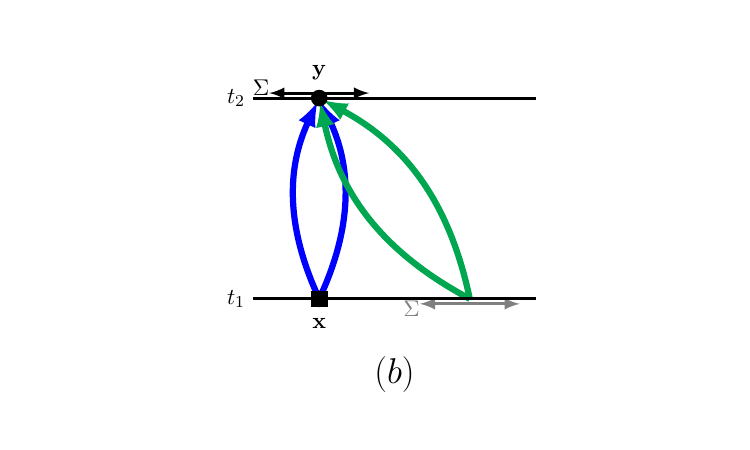}
\hspace{-3.8cm}\includegraphics[width=7.6cm,page=7]{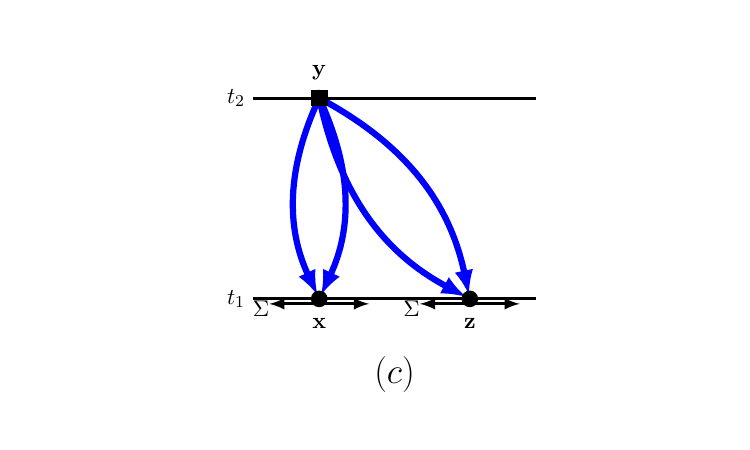}
\end{center}
\vskip-1.cm
\caption{\label{fig:examplediagram}Possible methods of computation for the $4 \times$ connected diagram of $C_{2 5}(t)$. Quark propagators are represented by solid lines and can be computed using the one-end trick (green) or point-to-all propagators (blue).}
\end{figure}

\begin{figure}[htb]
\begin{center}
\vskip-3.5cm
\includegraphics[width=13cm,page=1]{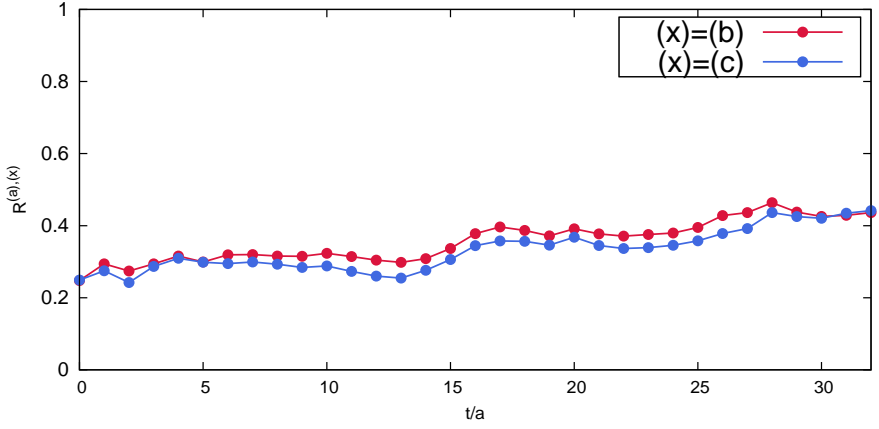}
\end{center}
\vskip-1.7cm
\caption{\label{fig:exampleperformance}Comparison of method~$(a)$ to method~$(b)$ and method~$(c)$ shown in Figure~\protect\ref{fig:examplediagram}.}
\end{figure}

As an example we briefly discuss a specific diagram contributing to $C_{2 5}(t)$, namely, the correlation between the $K \bar{K}$ molecule operator and the $K + \bar{K}$ scattering operator. Three possible methods of computation are sketched in Figure~\ref{fig:examplediagram}. Both ratios $R^{(a),(b)},R^{(a),(c)} < 1$ (cf.\ Figure~\ref{fig:exampleperformance}). Consequently, method~(a) (applying the one-end trick twice) is more efficient than method~(b) or (c). For a detailed discussion of all diagrams of the correlation matrix \eqref{EQN100} we refer to an upcoming publication.

\vskip.7cm

\begin{figure}[htb]
\begin{center}
\includegraphics[width=7.5cm,page=1]{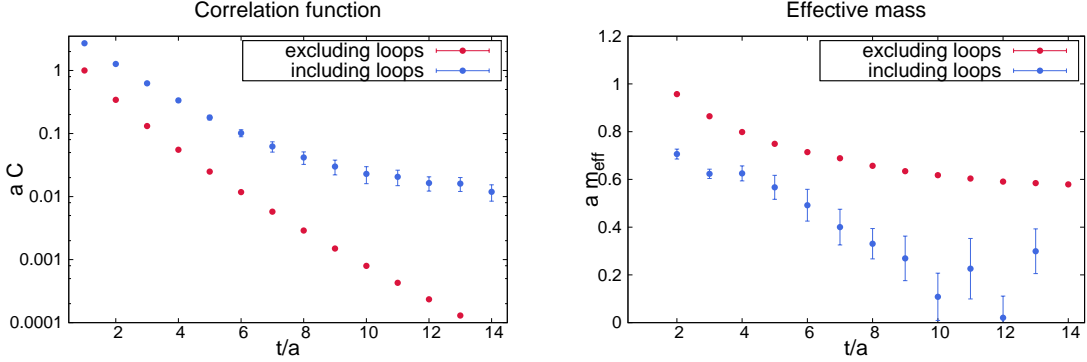}
\includegraphics[width=7.5cm,page=2]{GEPm}
\end{center}
\vskip-1.cm
\caption{\label{fig:effectofloops}$C_{4 4}(t)$ and the corresponding effective mass with (blue) and without (red) considering contributions from closed quark loops.}
\end{figure}

Finding the optimal method of computation is of particular importance for diagrams containing closed quark loops, i.e.\ diagrams, where quarks are created and annihilated within the same timeslice. Such diagrams are inherently noisy with relative statistical errors increasing exponentially with respect to the temporal separation $t$. On the other hand, diagrams with closed quark loops are crucial for our study of $a_0(980)$, because they lead to a non-vanishing correlation of two-quark and four-quark operators, i.e.\ they allow for $s \bar{s}$-pairs creation and annihilation. Moreover, even for correlators between two four-quark operators their contribution is sizeable and cannot be neglected as it has been done in the past (cf.\ e.g.\ \cite{Prelovsek:2010kg,Alexandrou:2012rm}). This is demonstrated in Figure~\ref{fig:effectofloops}, where we show $C_{4 4}(t)$ and the corresponding effective mass both with and without closed quark loops taken into account. Similar observations have been reported in \cite{Wakayama:2014gpa}.

\section{\label{sec:results}Physical results}

We determine effective energies $E_n^\textrm{eff}$ and corresponding eigenvectors $\mathbf{v}_n$ by solving the standard generalized eigenvalue problem (GEP)
\begin{eqnarray}
C(t) \mathbf{v}_n(t,t_r) = \lambda_n(t,t_r) C(t_r) \mathbf{v}_n(t,t_r) \quad , \quad E_n \overset{t \gg 1}{=} E_n^\textrm{eff}(t,t_r) = \frac{1}{a} \ln\bigg(\frac{\lambda_n(t,t_r)}{\lambda_n(t+a, t_r)}\bigg)
\end{eqnarray}
with $t_r = a$. We also apply a complementary approch, the AMIAS method, where multi-exponential fits are performed with initial conditions chosen by a measure proportional to $e^{-\chi^2/2}$, where $\chi^2$ is the chi-squared of the corresponding fits.. The AMIAS method might provide several advantages compared to solving the GEP, in particular it is possible to omit a number of very noisy correlation matrix elements in the analysis. For details cf.\ \cite{TALK_JACOB,Alexandrou:2014mka}.

\begin{figure}[htb]
\begin{center}
\includegraphics[width=14cm,page=1]{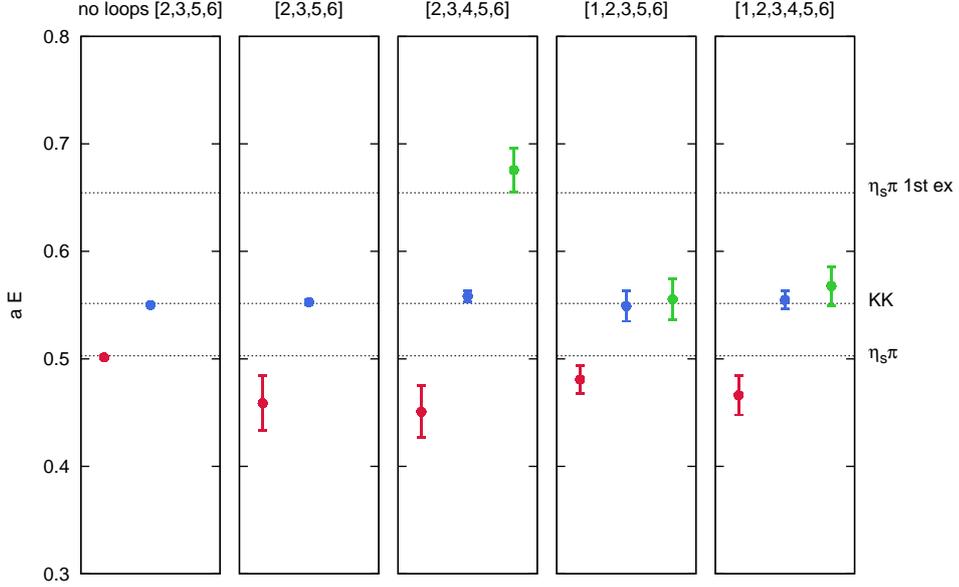}
\end{center}
\vskip-2cm
\caption{\label{fig:energycolumns}Lowest energy levels from several GEP analyses using operators according to the indices listed in the caption of each column. Note that an additional low-lying state around the two-particle thresholds is obtained only after including the quark-antiquark operator $\mathcal{O}^{q \bar q}$ (index $1$).}
\end{figure}

The lowest two or three energy levels extracted from several submatrices of the correlation matrix \eqref{EQN100} are shown in Figure~\ref{fig:energycolumns}. The indices listed in the caption of each column indicate which subset of operators of \eqref{eq:operatorone} to \eqref{eq:operatorsix} have been considered in the corresponding submatrix. The two-particle thresholds of $\eta + \pi$ and $K + \bar{K}$, as well as the first momentum excitation of $\eta + \pi$ are represented by dotted black lines.

In the first column we neglect closed quark loops and, we obtain rather precise results for two states consistent with the two particle thresholds (this confirms our previous results in \cite{Alexandrou:2012rm}). In all other columns, instead, we take closed quark loops into account. Note that an additional low-lying state around the two-particle thresholds is obtained only after including the quark-antiquark operator $\mathcal{O}^{q \bar q}$.

\begin{figure}[htb]
\begin{center}
\includegraphics[width=4cm,page=4]{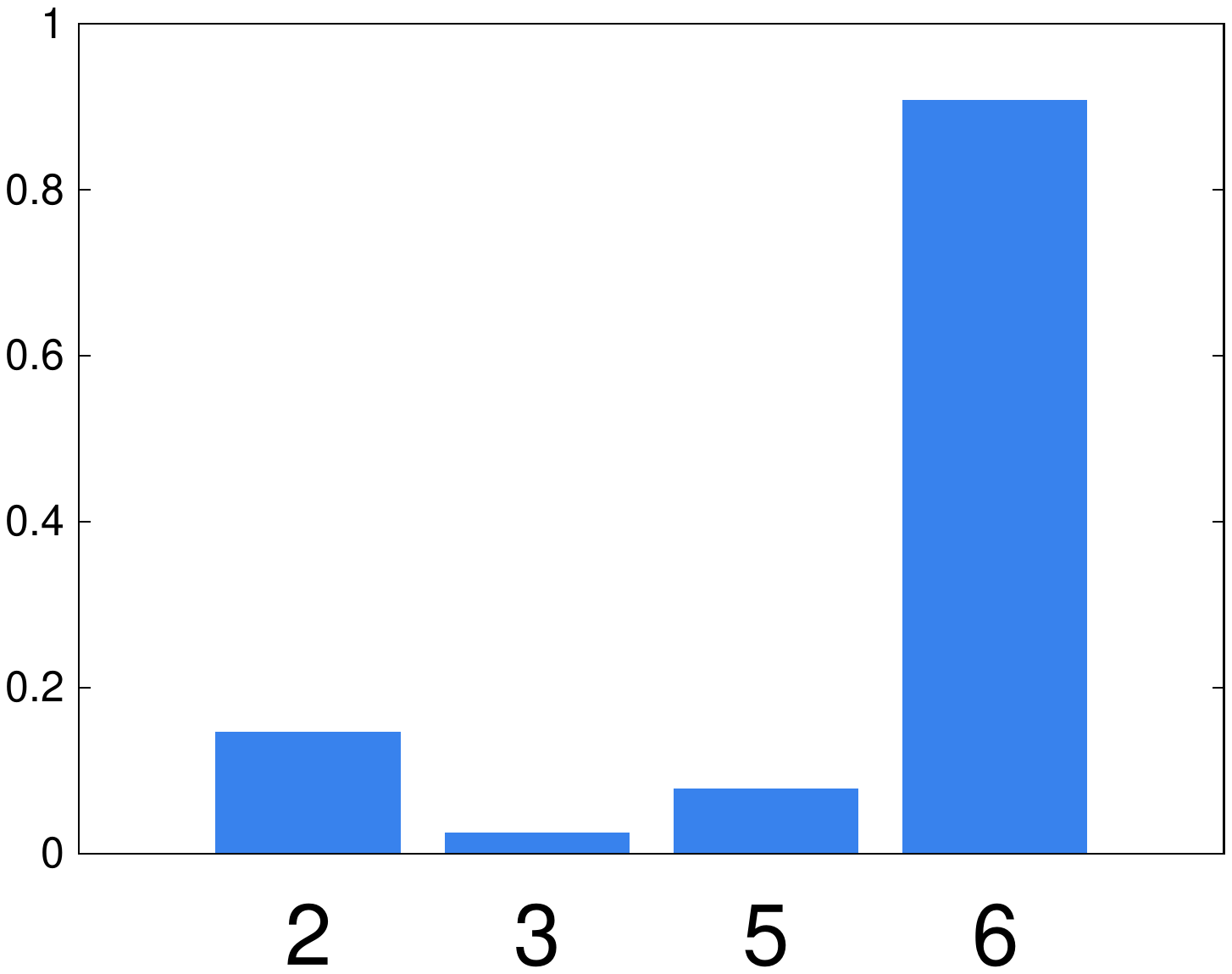}
\includegraphics[width=4cm,page=3]{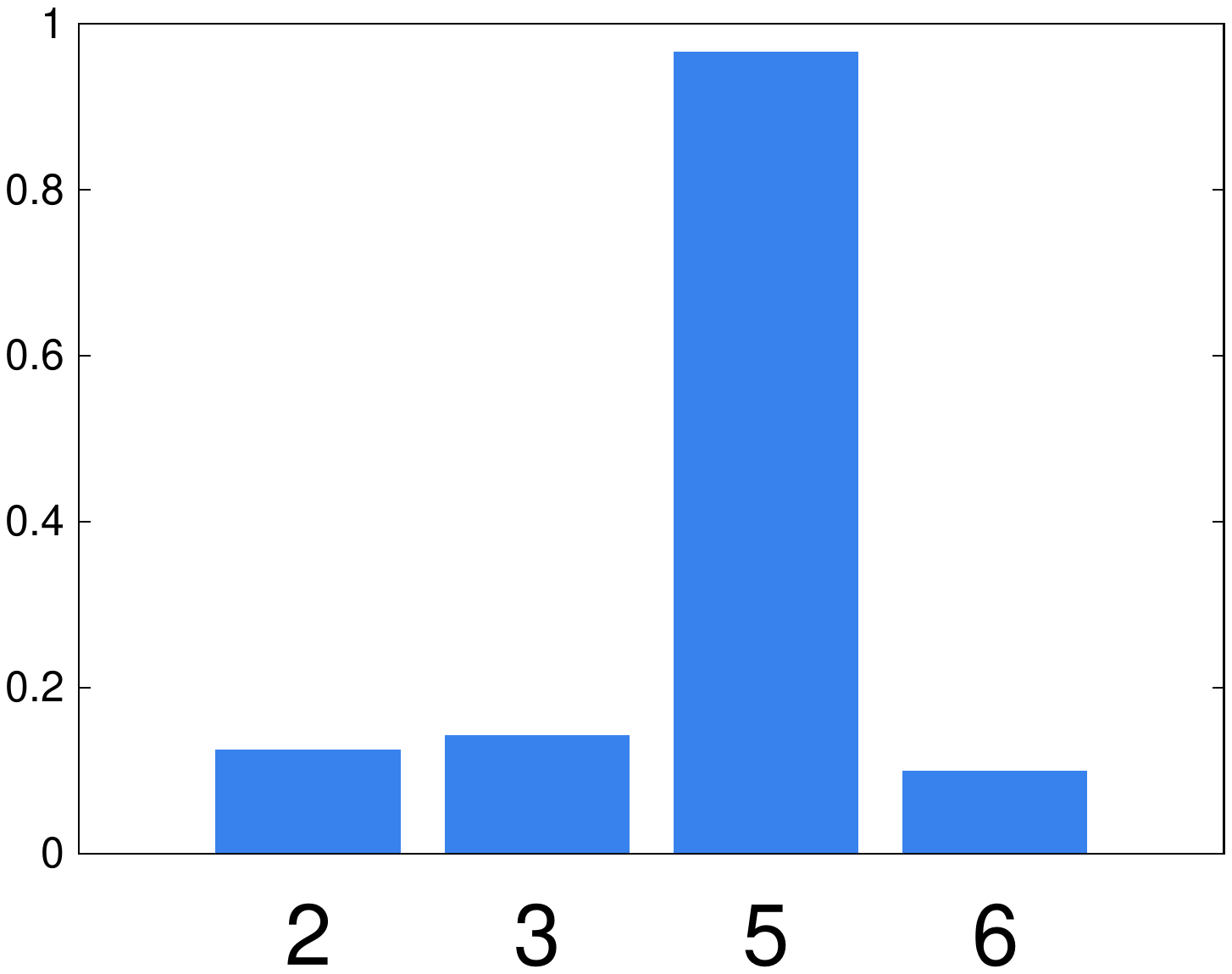} \\
\includegraphics[width=4cm,page=9]{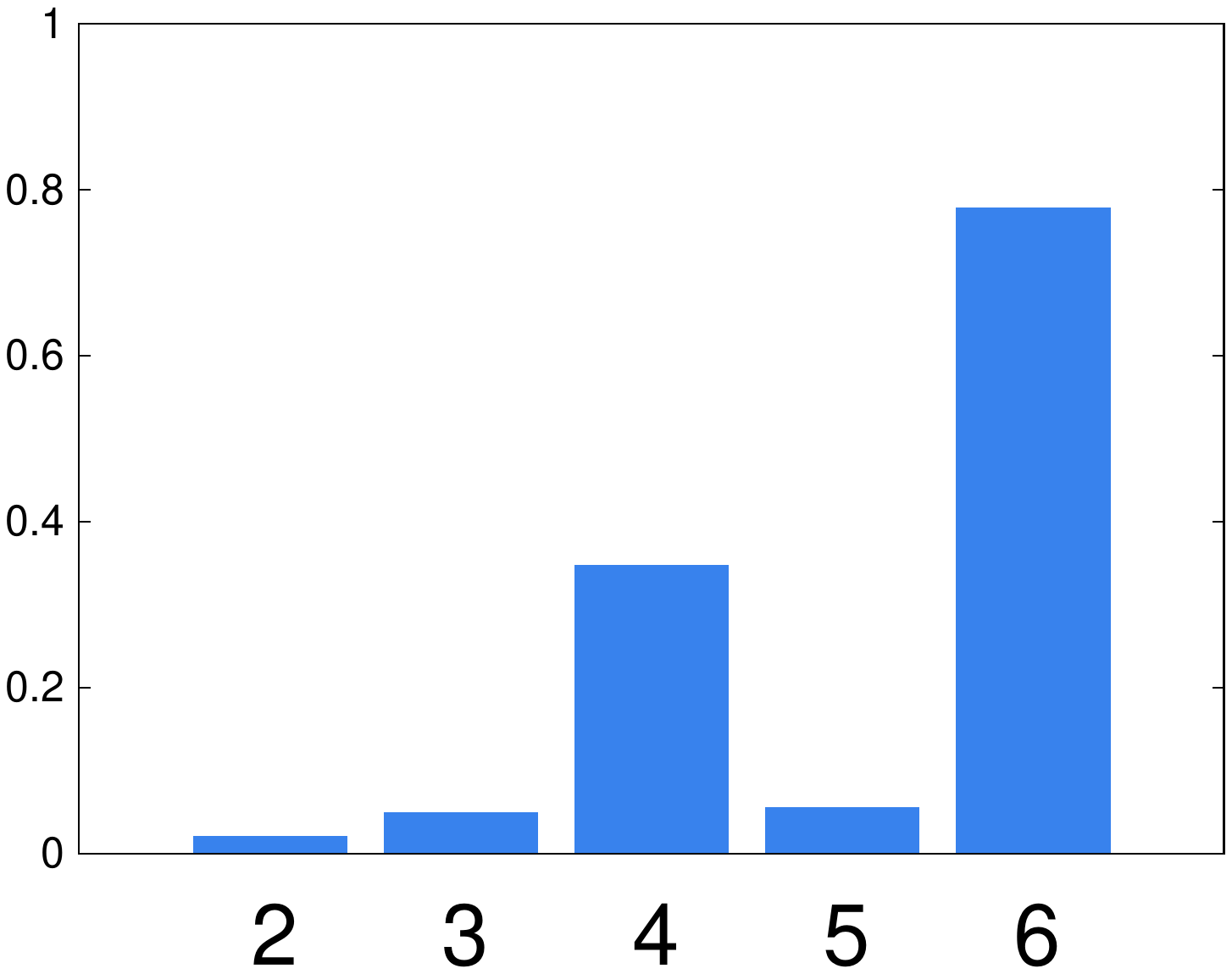}
\includegraphics[width=4cm,page=8]{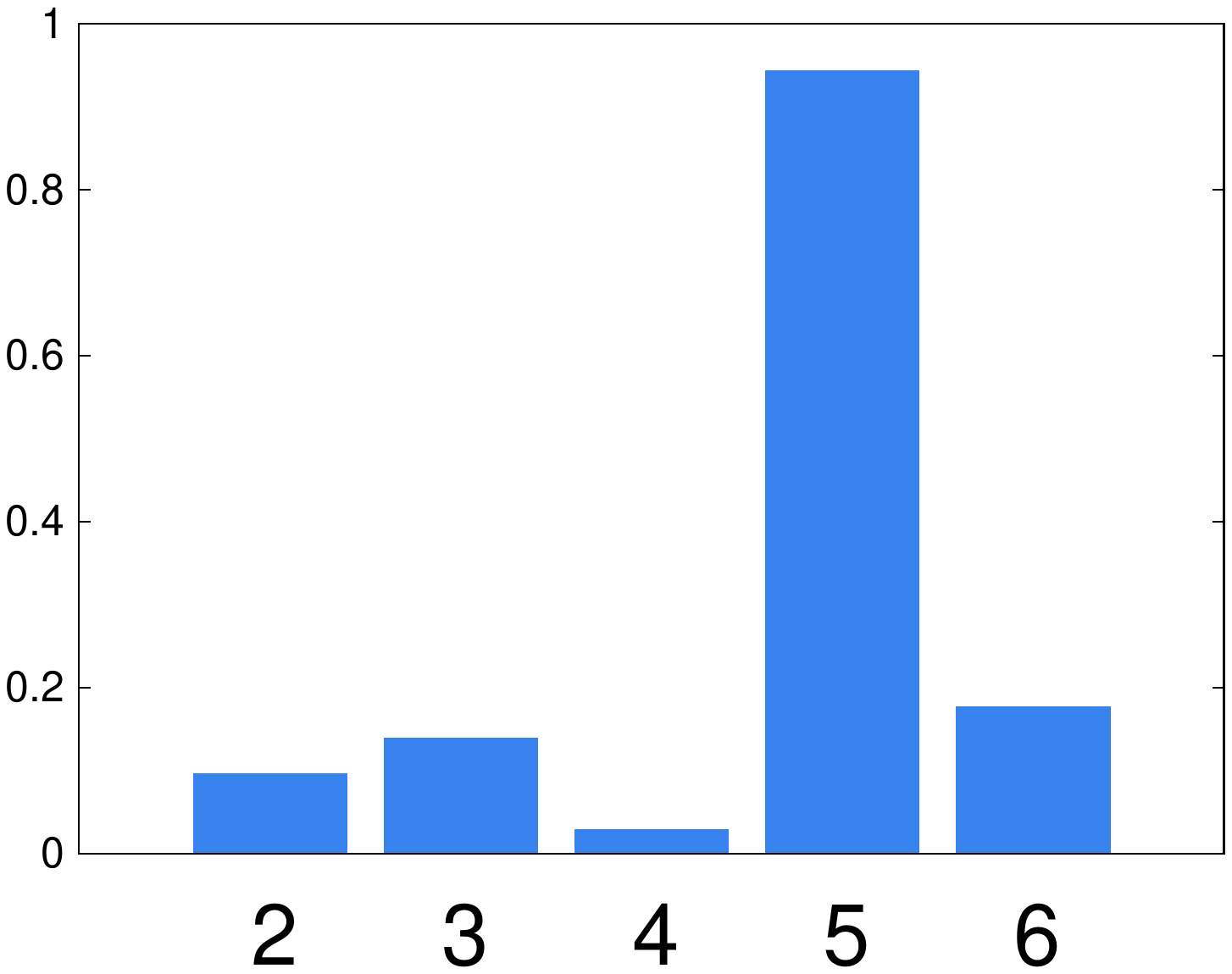}
\includegraphics[width=4cm,page=7]{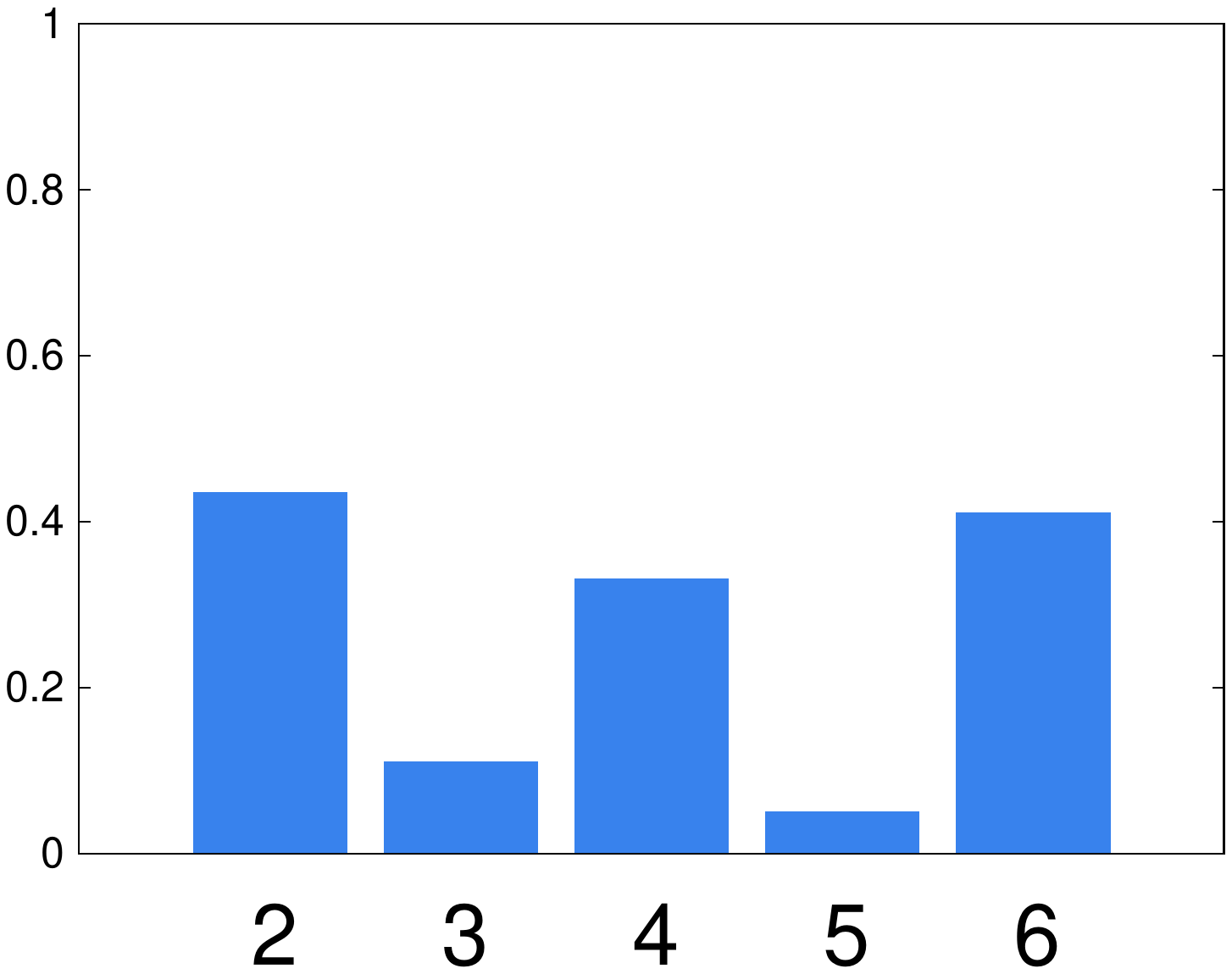} \\
\includegraphics[width=4cm,page=14]{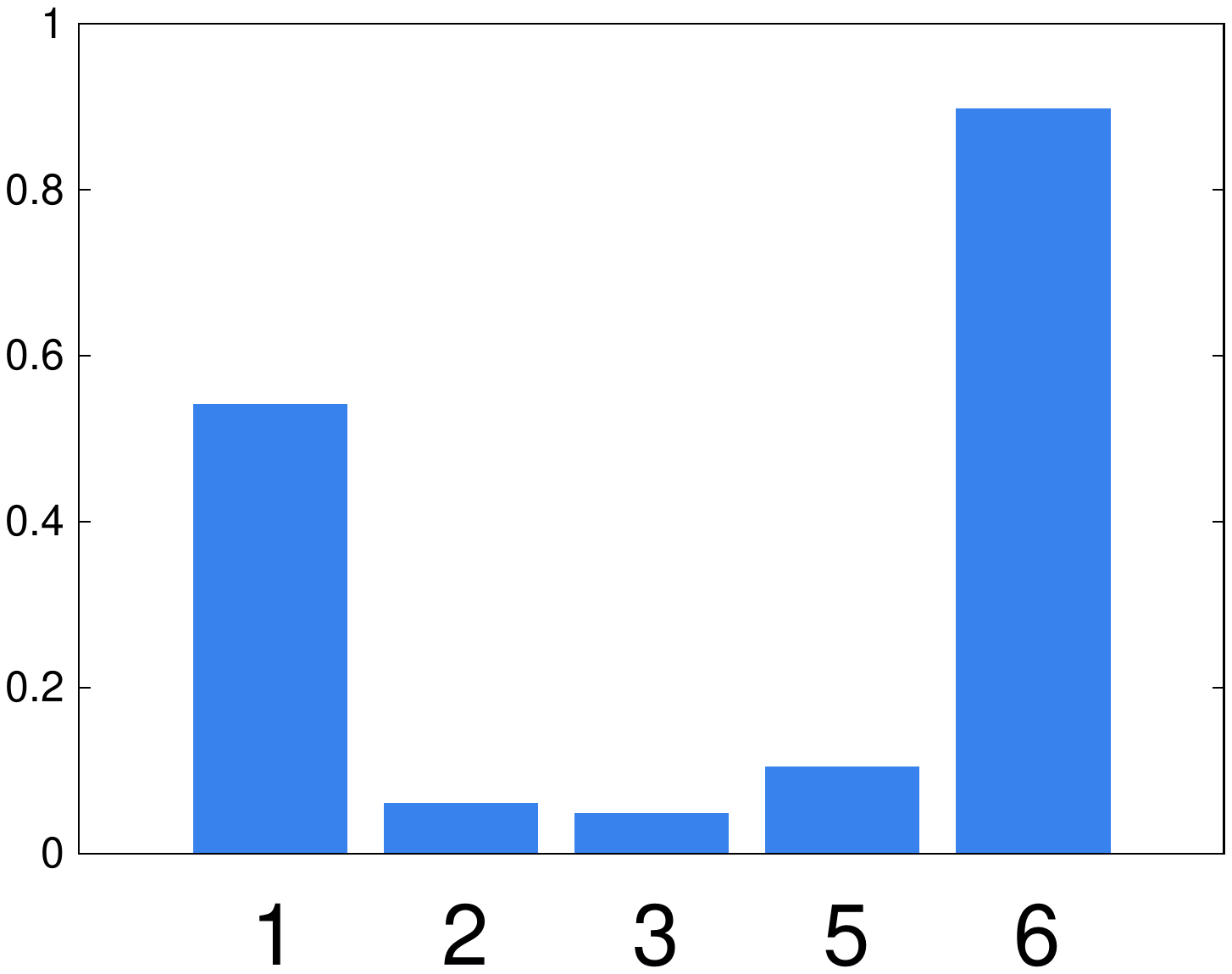}
\includegraphics[width=4cm,page=13]{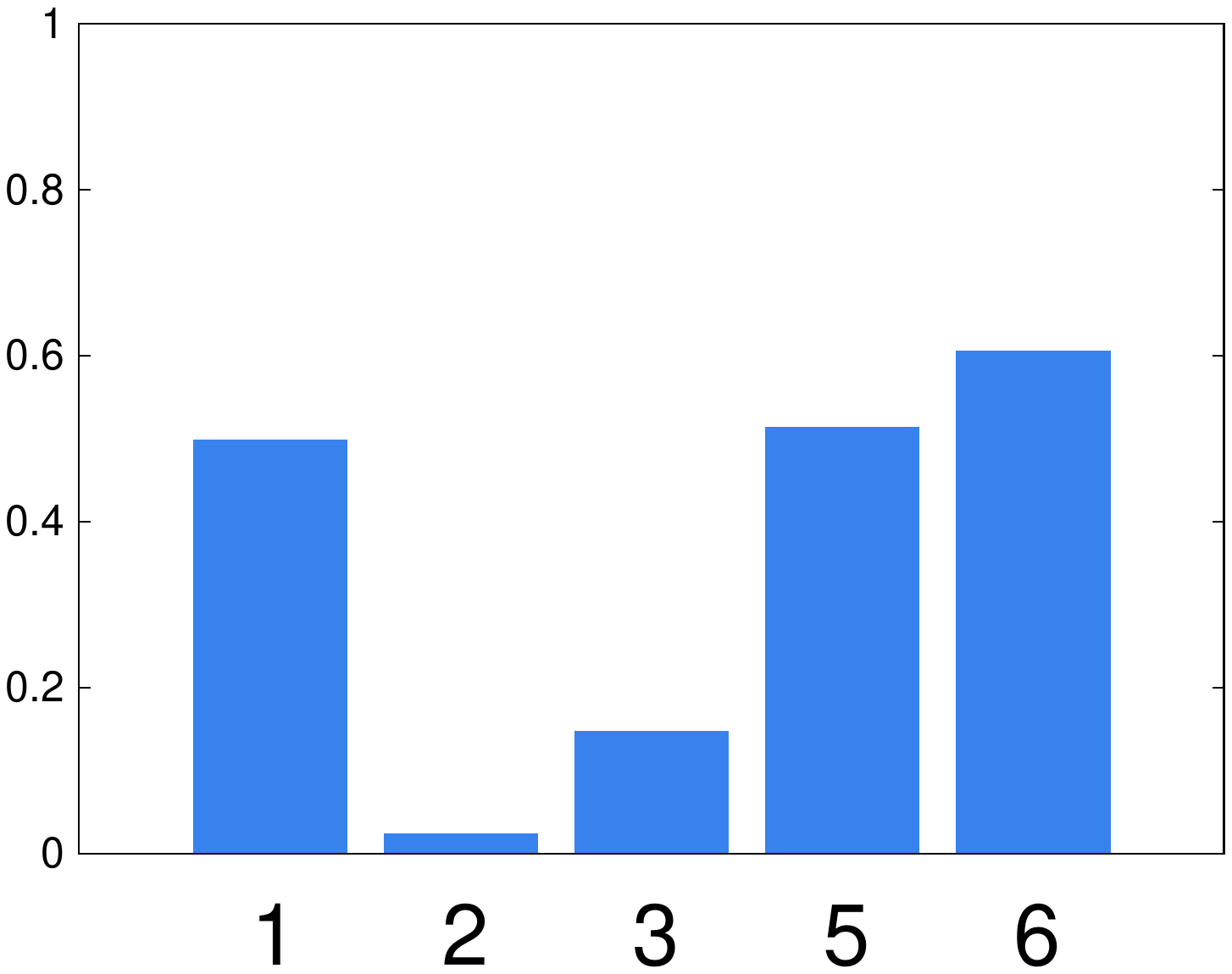}
\includegraphics[width=4cm,page=12]{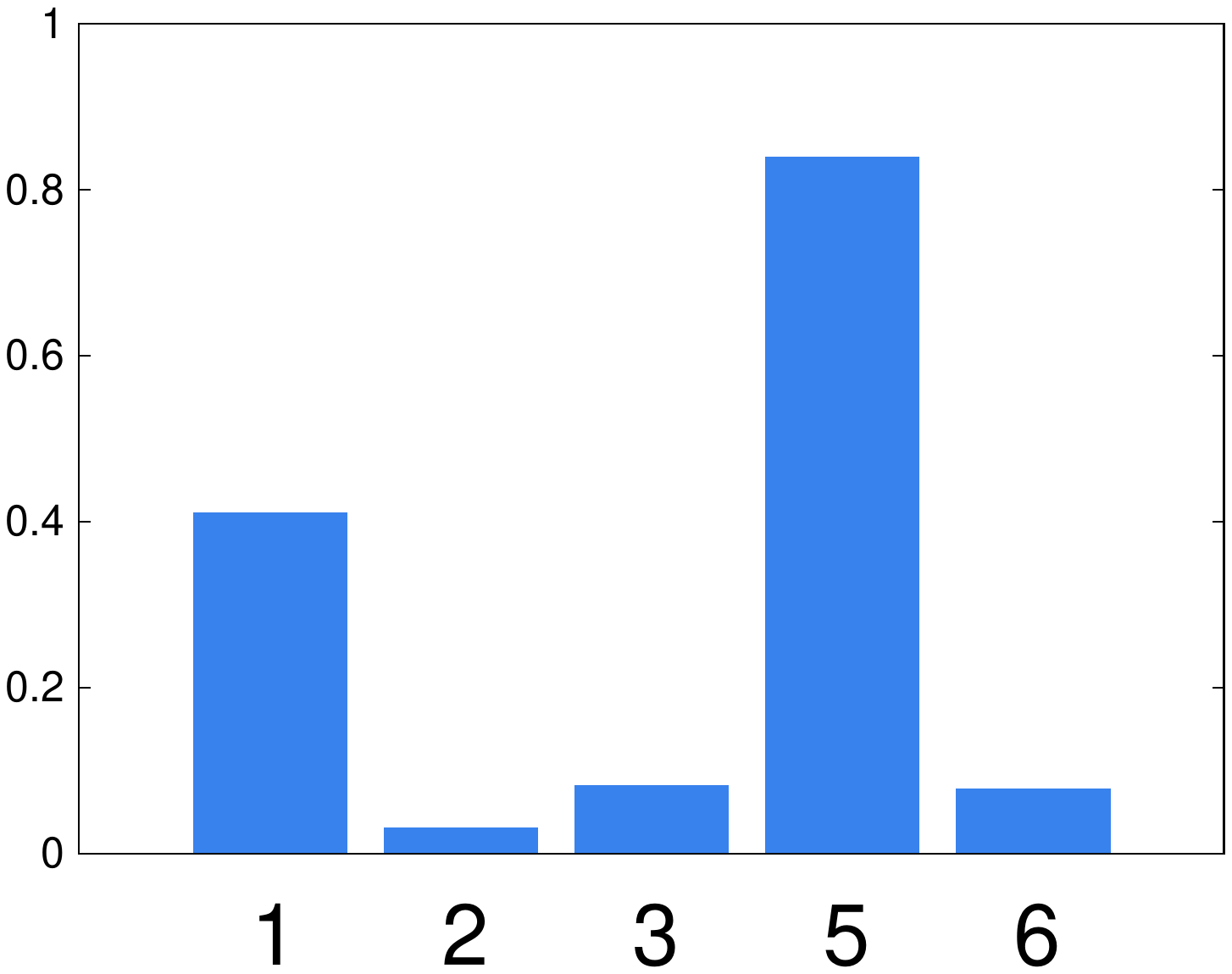} \\
\includegraphics[width=4cm,page=20]{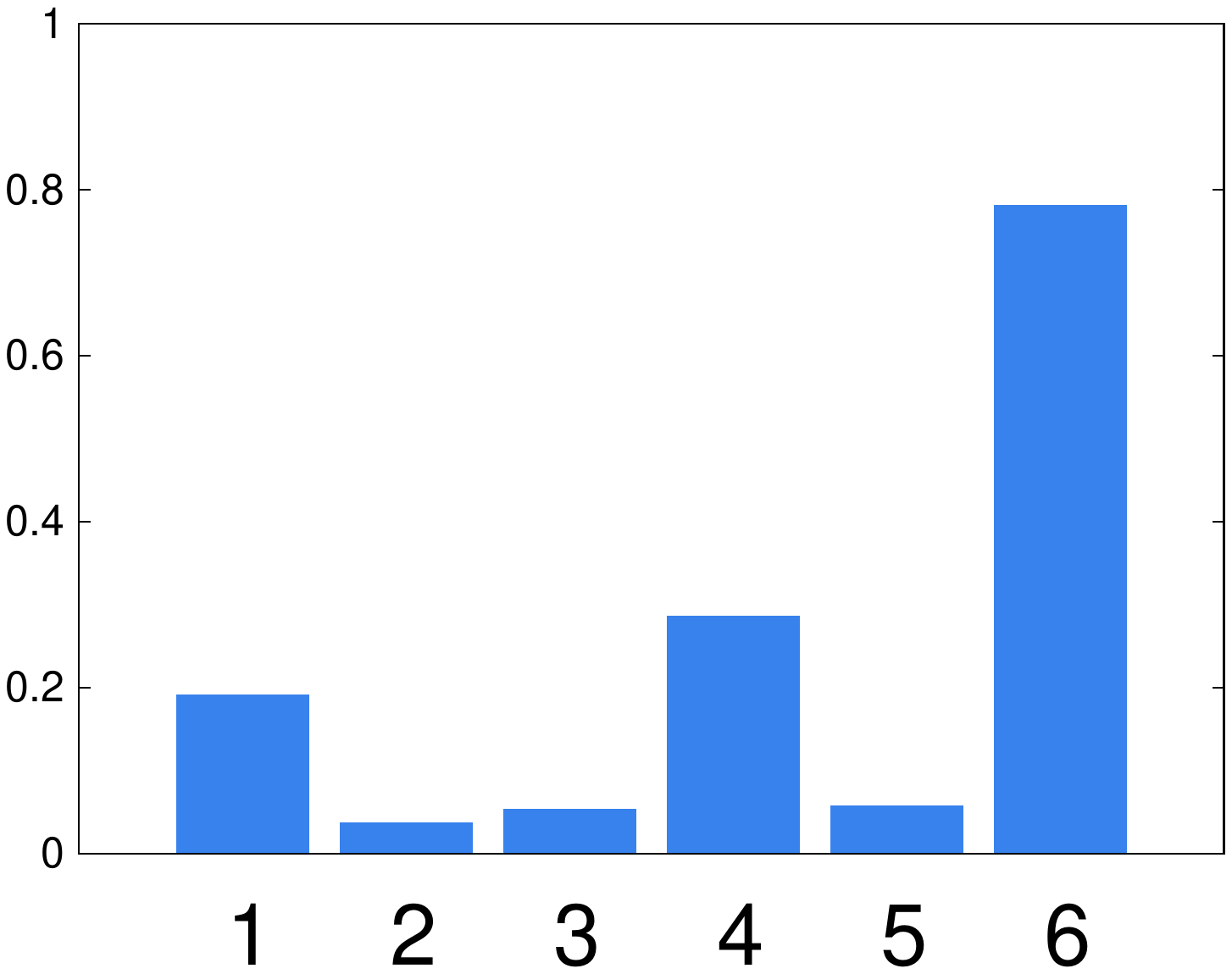}
\includegraphics[width=4cm,page=19]{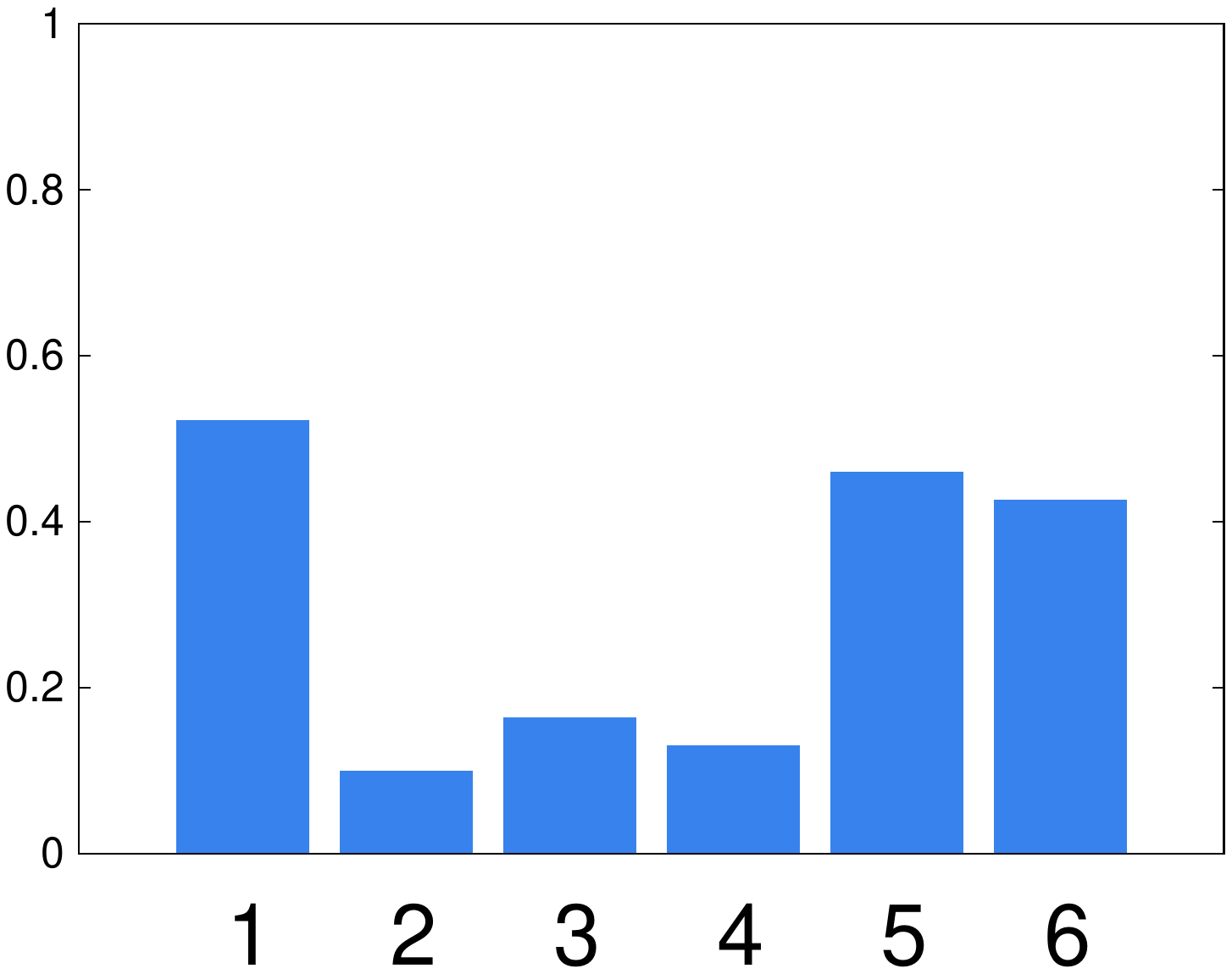}
\includegraphics[width=4cm,page=18]{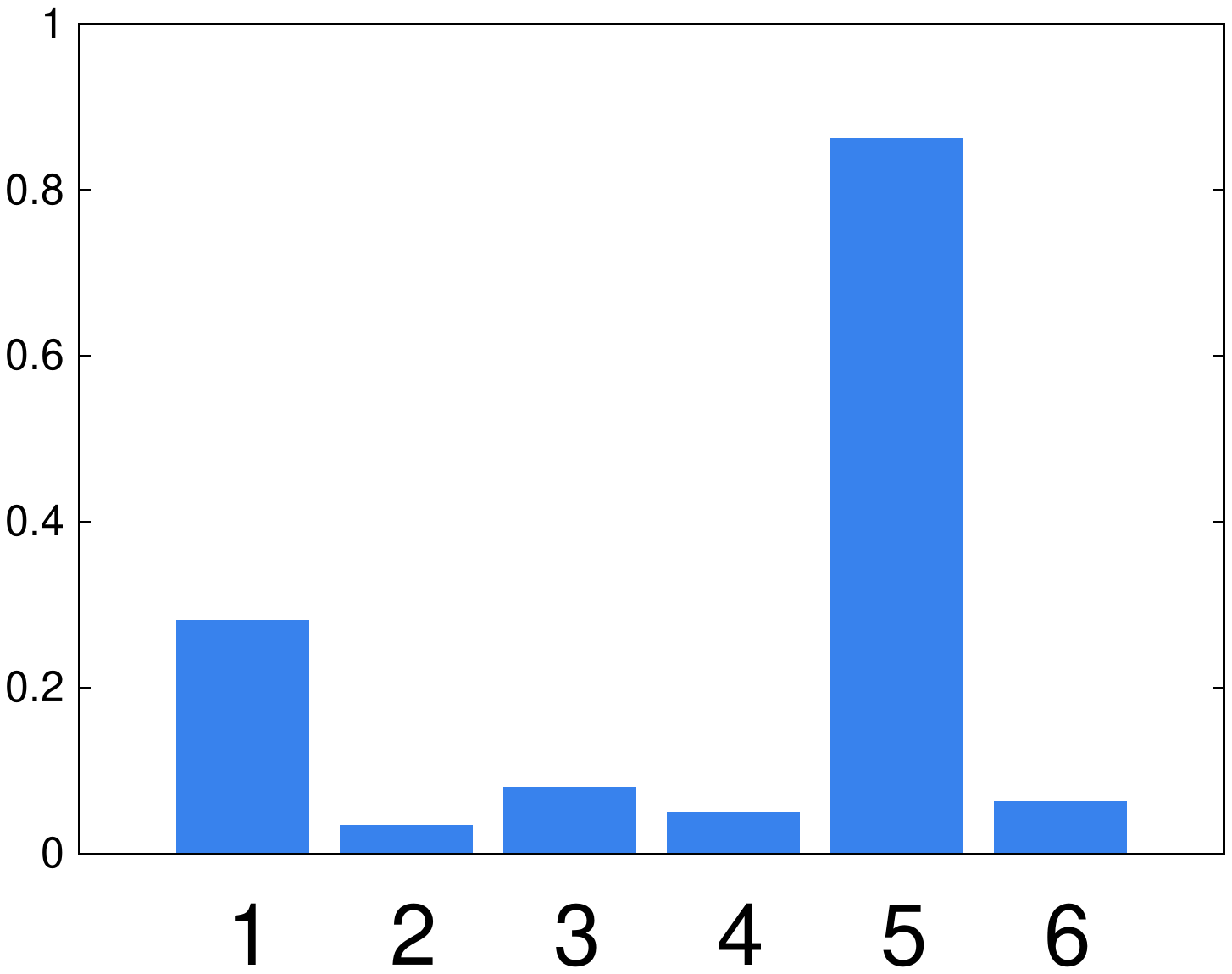}
\end{center}
\vskip-.5cm
\caption{ \label{fig:evs}Eigenvector components corresponding to the analyses with closed quark loops from Figure~\protect\ref{fig:energycolumns}. The first/second/third/fourth row of eigenvector components corresponds to the energies of the second/third/fourth/fifth column in Figure~\protect\ref{fig:energycolumns}. In each row the eigenvectors, i.e.\ the subplots, are ordered from left to right according to increasing energy.}
\end{figure}

The eigenvector components corresponding to these analyses are shown in Figure~\ref{fig:evs}. Note, in particular, that the analyses with operators [1,2,3,5,6] and [1,2,3,4,5,6] (the two rightmost columns in Figure~\ref{fig:energycolumns}), where an additional low-lying state has been found, are dominated by the same three operators: quark-antiquark (index 1), and $K+ \bar{K}$ and $\eta_s + \pi$ scattering (indices 5 and 6). We interpret these results as an indication that the $a_0(980)$ is not predominantly a tetraquark state, but rather has a significant quark-antiquark component.

In this context we refer again to the $a_0(980)$ study reported in \cite{Dudek:2016cru}, where a resonance is clearly identified as a pole near the $K\bar{K}$ threshold with strong coupling to both $K + \bar K$ and $\eta + \pi$ channels. 

\begin{acknowledgments}

J.B.\ and M.W.\ acknowledge support by the Emmy Noether Programme of the DFG (German Research Foundation), grant WA 3000/1-1.
The work of M.G.\ was supported by the European Commission, European Social Fund and Calabria Region, that disclaim any liability for the use that can be done of the information provided in this paper.

This work was supported in part by the Helmholtz International Center for FAIR within the framework of the LOEWE program launched by the State of Hesse.

Computations have been performed using the Chroma software library \cite{Edwards:2004sx}.

Calculations on the LOEWE-CSC high-performance computer of Johann Wolfgang Goethe-University Frankfurt am Main were conducted for this research. We would like to thank HPC-Hessen, funded by the State Ministry of Higher Education, Research and the Arts, for programming advice.

\end{acknowledgments}



\end{document}